# Remindful: Designing Reminder Systems for Caregiver Interpretation in Dementia Care

Caregiver Interpretation in Dementia Reminder Systems

## JOY LAI

Institution of Biomedical Engineering, University of Toronto, joy.lai@mail.utoronto.ca

## ALEX MIHAILIDIS

Institution of Biomedical Engineering, University of Toronto, alex.mihailidis@utoronto.ca

Digital reminder systems are widely used in dementia care to support everyday tasks, but they are typically designed for one-way prompting rather than helping caregivers interpret engagement over time. We present Remindful, a caregiver-informed reminder platform that extends task prompting with caregiver-facing alerts, summaries, and review features to support awareness in home-based dementia care. Drawing on formative caregiver interviews, lived-experience advisor input, and in-home deployments with two caregiver-PLwD dyads, we examine how reminder-based caregiver awareness functions in practice. Our findings show that reminder systems can support caregiver reassurance, household coordination, and awareness of routines over time, but that reminder interaction data is highly context-dependent. Household participation, prompt attribution, routine mismatch, accessibility barriers, and technical failures all shaped what reminder logs could reasonably mean. We argue that reminder systems should not be treated as neutral behavioral sensors, but designed as assistive infrastructures for caregiver interpretation that preserve uncertainty and support contextual sensemaking in real homes.

CCS CONCEPTS • **Human-centered computing~Human computer interaction (HCI)** • **Human-centered computing~Human computer interaction (HCI)~Interactive systems and tools** • Human-centered computing~Accessibility~Accessibility technologies

## 1 INTRODUCTION

People living with dementia (PLwD) and their caregivers face ongoing challenges in maintaining everyday routines at home, including medications, meals, hygiene, and appointments [20]. Digital reminder systems are a common form of assistive technology that supports these routines, prompting task completion and reducing the need for repeated verbal reminders [9, 16, 17, 21]. However, most reminder systems remain primarily task-focused: they deliver prompts, but offer limited support for verification and understanding how engagement with those routines changes over time [16, 21].

For caregivers, workload is often tied not only to single missed tasks, but to uncertainty about whether routines are being completed consistently and whether changes in those routines may signal evolving support needs [5, 7]. In this sense, reminder systems may have value beyond prompting alone. Because they are already embedded in everyday care routines, they create structured interaction traces such as missed reminders, delayed responses, and follow-up interactions that could potentially support caregiver awareness over time. This could encompass awareness of missed tasks, broader behavioral changes, and emerging patterns that may matter within a given household.

At the same time, extending reminder systems into caregiver-support tools is not straightforward. Reminder interactions are not neutral records of behaviour. A missed or delayed response may reflect meaningful change, but it may also reflect a timing mismatch, difficulty engaging with the prompt, shifting household routines, or technical problems. In dementia care, these systems are rarely used by a single, isolated person. Caregiving technologies are often configured, interpreted, and socially negotiated within the home, where issues of surveillance, autonomy, and shared responsibility shape how systems are used [3, 13, 15, 18, 22].

In this paper, we present Remindful, a reminder-based assistive system designed to extend prompting with caregiver-facing awareness and review support. The broader goal of this work is to examine whether reminder systems, which are already familiar and relatively low-burden assistive tools, can be extended to provide caregivers with useful feedback on routine changes over time. Our approach was informed by formative interviews with caregivers, input from lived-experience advisors, and in-home deployments with caregiver-PLwD dyads [10]. Rather than evaluating reminder systems as simple adherence tools, we examine how reminder-based awareness is produced, interpreted, and used in practice.

This paper makes three contributions. First, it presents Remindful, a caregiver-informed reminder system that extends task prompting with caregiver-facing alerts, summaries, and review features for home-based dementia care. Second, through in-home deployments, it shows how reminder systems can support caregiver awareness through reassurance, coordination, and routine tracking, even when use is distributed across households rather than attributable to one individual alone. Third, it identifies key interpretive challenges that shape what reminder interaction data can mean in practice, including household context, prompt attribution, routine mismatch, accessibility barriers, and technical reliability. Together, these findings reframe reminder systems not as neutral adherence trackers, but as assistive infrastructures that must be designed around interpretation and uncertainty.

## 2 BACKGROUND

Digital reminder systems are a well-established form of assistive technology for supporting daily routines in dementia care, including medications, meals, hygiene, and appointments. Prior work shows that these systems can reduce repetitive prompting burden and support routine adherence, but they typically remain focused on task initiation rather than helping caregivers understand how engagement changes over time [9, 16, 17, 21]. Even when reminder systems include some form of logging or remote oversight, they generally provide limited support for interpreting broader patterns of engagement or for helping caregivers verify what actually happened in context [16, 23].

A related body of work has explored anomaly detection and behavioral monitoring in dementia care. This literature has largely focused on sensor-rich approaches, including wearables, cameras, GPS, environmental sensors, and smart home systems [12, 28]. These approaches can provide continuous monitoring and timely alerts, but they also entail recurring trade-offs regarding privacy, user compliance, false positives, comfort, cost, and installation burden [3, 21, 28]. Many systems in this space have also been developed and evaluated in controlled, simulated, or otherwise limited settings, with less evidence about long-term use and integration into everyday caregiving routines [12].

This gap makes reminder systems an important design opportunity. Because they are already used in dementia care and operate on familiar devices such as tablets or smartphones, they may offer a lower-friction and potentially more privacy-preserving foundation for caregiver support than adding a separate monitoring infrastructure [9, 17, 21]. At the same time, prior work suggests that reminder interactions may contain useful behavioral signals, but that any such signals must be interpreted carefully and in context rather than treated as direct evidence of behavior [16, 23].

Research on caregiver-facing verification and human-AI collaboration suggests that the core challenge is not simply detecting whether something unusual occurred, but supporting caregiver interpretation in ways that are understandable, low burden, and adjustable. Existing verification strategies still often rely on direct observation, follow-up calls, or sensor-based inference [4, 8, 19, 24, 26]. Work on explainable and mixed-initiative AI similarly shows that people are more likely to rely appropriately on system outputs when they can understand, inspect, and override them [1, 6, 14, 25]. In caregiving contexts, this suggests that support systems should help caregivers build and assess trustworthy evidence rather than simply present opaque judgments [2, 3].

This perspective is especially important for accessibility research. In reminder-based caregiver support, accessibility is not only about whether a prompt can be delivered, but whether the resulting interaction and resulting data remain meaningful, interpretable, and usable in context. HCI scholarship on dementia technologies has emphasized that caregiving technologies are relational and emotionally interpreted within the home, and that systems intended to support safety or adherence can also introduce surveillance concerns, interpretation burden, or loss of control if they are poorly designed [3, 13, 15, 18, 22]. Against this background, the present paper is neither a clinical prediction study nor a pure anomaly-detection paper. Instead, it examines reminder systems as assistive technologies that might be extended to support caregiver awareness in real homes. The focus is on whether reminder-based awareness is meaningful, trustworthy, and usable in practice, and on what kinds of interpretive, contextual, and accessibility challenges arise when caregivers are asked to rely on reminder interaction data rather than simply use reminders as one-way prompting tools.

## 3 SYSTEM AND STUDY OVERVIEW

### 3.1 Caregiver-informed Design Grounding

Remindful was grounded in a formative qualitative phase designed to identify which routine changes caregivers considered meaningful, which forms of system feedback they would find useful, and the constraints these expectations posed for reminder-based assistive technologies. Semi-structured interviews were conducted with 10 caregivers, with optional participation from PLwD limited to dyadic interviews when feasible; one PLwD participated in this way to provide a complementary perspective on interaction preferences and system framing. Across interviews, caregivers described priorities around verifying whether tasks were actually completed, receiving summaries rather than raw data, reducing monitoring burden, and retaining caregiver control. These priorities informed later system requirements and deployment decisions.

This grounding was strengthened through sustained collaboration with two lived-experience advisors from the EPLED program, who participated as methodological partners rather than as study participants [27]. Their involvement extended beyond simple feedback: they helped shape interview framing, surface ethical and practical blind spots, and support interpretation so that design assumptions remained aligned with real caregiving experience [11].

## 3.2 The Remindful System

Remindful was implemented as a distributed reminder platform with two connected interfaces: an always-on in-home tablet system for the PLwD, and a caregiver-facing mobile app used to configure reminders, monitor activity, and review reports. Tablets were placed in meaningful locations within the home, such as the kitchen, bedroom, or living room, and reminders were assigned to one or more of these locations so that prompts could be aligned with the activity being supported. This location-based design, shown in Figure 1, treated physical space as part of the interaction model rather than as background context.

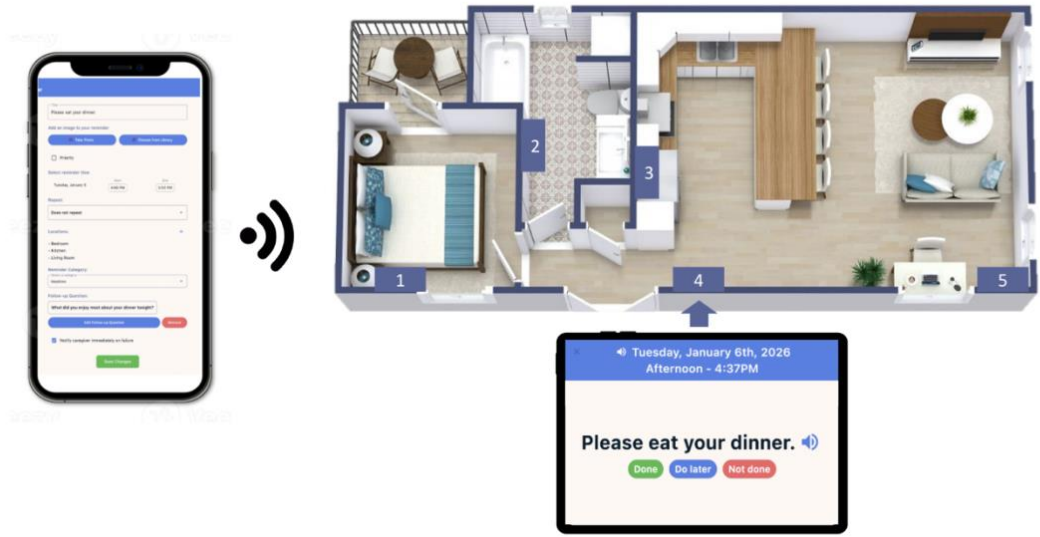

Figure 1. Example of the Remindful reminder set-up in the home. Floorplan image adapted from RoomSketcher. Caregiver and PLwD interface screenshots are authors' own.

Using the caregiver app, caregivers could create reminders, assign recurrence schedules and priority levels, specify display locations, and optionally attach follow-up questions. When a reminder appeared on an in-home iPad, the PLwD could respond using one of three explicit options: Done, Do Later, or Not Done, as shown in Figure 2. If a reminder was marked Done, the system could present an optional follow-up question. Follow-up responses could be entered by text or by voice after an explicit tap on the microphone icon; passive listening was not used. Each interaction generated structured log data including reminder identity, scheduled time, display location, response type, response timestamp, and any follow-up response.

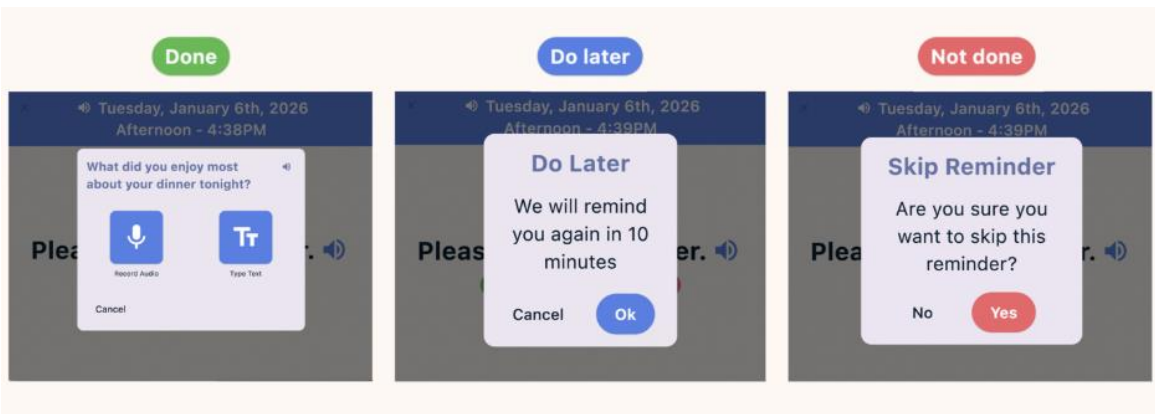

Figure 2: Example PLwD-facing reminder interactions in Remindful.

The deployed system also included caregiver-facing reports and alerts, shown in Figure 3. Daily reports highlighted missed reminders and other items needing attention, while longer-term summaries aggregated interaction patterns over time, by location, and by reminder type. In the home study, daily reports and alerts were active within the system, and caregivers could rate alerts and provide feedback through the interface. Longer-term summaries were generated from reminder interaction data and uploaded weekly to the app by the research team during deployment. These reports were descriptive rather than diagnostic: they summarized interaction patterns such as acknowledgment rates, location effects, timing trends, and repeated reminders, rather than making automated claims about decline or risk.

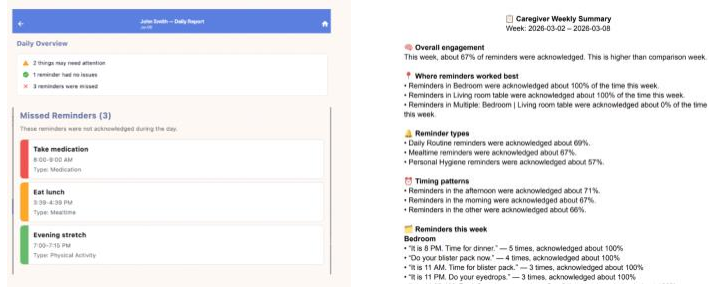

Figure 3: Example sections of the daily (left) and long-term (right) reports.

At the architectural level, shown in Figure 4, Remindful drew on a broader design structure spanning detection, verification, and presentation. However, the deployment examined in this paper focused specifically on the prompting, logging, alerting, and reporting components that households and caregivers interacted with in practice. Rather than evaluating autonomous anomaly detection performance, this study examined whether reminder-based outputs were meaningful, trustworthy, and workable within everyday care routines.

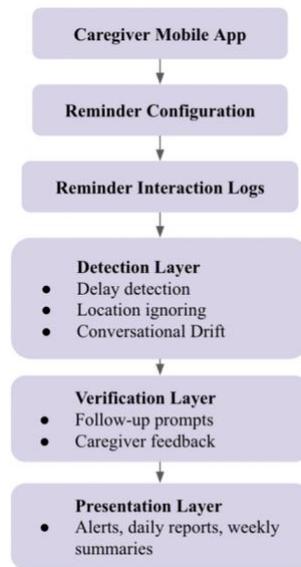

Figure 4: Diagram of Remindful's architectural pipeline

### 3.3 In-home Deployment Study

We conducted an in-home deployment study with two caregiver-PLwD dyads to examine how Remindful functioned in practice. The goal of this phase was not to evaluate clinical effectiveness or predictive accuracy, but to understand interaction reliability, caregiver interpretation, workflow fit, and the contextual factors that determined whether reminder-based outputs were meaningful in real homes. The deployment therefore focused on a human-centered assessment of the integrated reminder system, including reminder delivery, caregiver configuration, alerts, summaries, and follow-up workflows.

Data collected during deployment included reminder interaction logs, caregiver feedback and usability ratings, post-deployment interviews, and open-ended reports on trust, burden, usefulness, and system fit within everyday care routines. Because the study was deployment-oriented and the sample was small, the analysis was descriptive rather than comparative or inferential. The purpose was to identify recurring use patterns, interpretation challenges, and design implications rather than to test outcome hypotheses.

Analysis combined reminder interaction logs with post-deployment interviews, caregiver feedback, and usability responses to build a descriptive account of how the system functioned in practice. We used a descriptive thematic analysis to identify recurring patterns in use, interpretation, and breakdown across the two homes. Themes were developed through iterative review of qualitative materials alongside deployment logs, with attention to how reported caregiver experiences aligned with or complicated the interaction record. The resulting themes reflect both reported caregiver experiences and observed interaction trends in the deployment data.

The two dyads represented meaningfully different caregiving arrangements, as shown in Table 1. The first dyad involved a granddaughter caregiver in a multigenerational private home and completed a 4-week deployment. Three in-home iPads were placed in the bedroom, kitchen, and basement, and 144 reminders were delivered during the study period. The second dyad involved a spouse caregiver living with the PLwD in an apartment, with additional remote family support from her brother, who did not live in the home. This deployment lasted for 8 weeks. Two in-home iPads were used for the PLwD, the spouse caregiver used a third iPad, and the remote caregiver used the mobile app on his phone; 979 reminders were delivered during this deployment. These contrasting contexts were analytically useful because they exposed different patterns of use, coordination, and interpretation. Across deployments, the core system remained the same, although small software refinements were made iteratively, including bug fixes and a later customization feature that allowed spoken reminders to use personalized phrasing such as the PLwD's name.

Table 1: Dyad context for deployment

| Dyad | Caregiver Relationship | Home setup | Deployment Duration | In-home devices | Reminders Delivered | Reminders Acknowledged |
|---|---|---|---|---|---|---|
| 1 | Granddaughter | Multigenerational private house | 4 weeks | 3 iPads across bedroom, kitchen, and basement. Caregiver used personal phone. | 144 | 120 |
| 2 | Spouse + remote family support | Apartment | 8 weeks | 2 PLwD iPads across bedroom and kitchen. Caregiver used 3$^{rd}$ iPad, secondary caregiver used personal phone. | 979 | 252 |

## 4 FINDINGS

The formative phase indicated that caregivers wanted reminder-based systems to do more than deliver prompts. Caregivers emphasized the value of summaries, support for verification, low-burden monitoring, and outputs they could interpret in context rather than streams of raw reminder events. The home deployments showed that these goals were only partially achievable in practice. Reminder interactions were shaped by household participation, prompt design, routine fit, and accessibility in ways that made the resulting data more complex than a simple record of task completion. Four themes summarize these findings.

### 4.1 Caregiver Awareness through Household Coordination

Across both deployments, the reminder system supported caregiver awareness not only by prompting the PLwD directly, but also by distributing attention and coordination across the home. Its value therefore extended beyond independent PLwD response to helping caregivers stay aware of routines, reduce uncertainty, and decide when to check in.

In Dyad 1, this was especially clear. The caregiver said the system had "kind of saved me the mental burden" of remembering appointments, visitors, and daily activities, and later described it as "a reminder for the whole house." In practice, reminders were often noticed and acted on by other household members as well, making the system useful as a shared coordination tool rather than only an individual prompt.

A related pattern appeared in Dyad 2, but with a different emphasis. There, the in-home caregiver explained that "it's a reminder for me, not for him," meaning the reminders often worked better as prompts for caregiver follow-up than as a direct self-management tool for the PLwD. Even when independent interaction was inconsistent, the system still helped structure oversight and reduced the need to remember everything manually.

These cases show that reminder systems may provide assistive value even when responses are socially distributed rather than attributable to one individual alone. In both homes, distributed use was part of how the system became useful.

## 4.2 Attribution and Relational Framing

Reminder effectiveness depended not only on what a prompt said, but on whether the PLwD recognized it as personally relevant and understood why the interaction was being asked of them. In both dyads, reminders worked better when they felt clearly directed and socially meaningful within the caregiving relationship. When that was missing, prompts could be ignored, experienced as burdensome, or become difficult to interpret.

In Dyad 1, attribution was a clear accessibility issue. Early reminders were less effective partly because the participant did not consistently recognize that the prompt was meant for him. After the spoken prompt was changed to include his name, the caregiver explained that "because it actually used his name, he was like, oh, like this is about me." Before that, she was unsure whether he thought the sound came from "the iPad or the TV or something." The problem was therefore not only whether the reminder content was understandable, but whether the system established self-relevance at the start of the interaction.

Dyad 2 revealed a related but distinct problem. There, confirmation prompts could become frustrating when they conflicted with the PLwD's own sense of what had already happened. The caregiver described recurring situations in which a reminder asked about a task he believed he had already completed. She explained that he would respond with comments such as, "Oh, it's so confusing. I'm telling you I did it and you said I didn't do it. I have no idea what I really do," and that the interaction could feel like "you forced me to do something." In these moments, the issue was not simple refusal or noncompliance. The reminder itself had become hard to reconcile with the participant's understanding of the routine.

At the same time, Dyad 2 showed that acceptance improved when the system was understood less as a neutral device demanding confirmation and more as part of an existing caregiving interaction. The caregiver explained that the reminders were helpful because hearing them prompted her to check whether tasks had been completed. In that framing, the system worked less by eliciting direct response from the PLwD and more by fitting into shared routine support already happening in the home.

These cases show that prompt attribution and relational framing shaped whether reminders were acceptable and usable in the first place. A prompt could fail if it was not perceived as personally directed, or be rejected if its role felt unclear or socially misaligned.

## 4.3 Ambiguity in Reminder Logs

A central finding across deployments was that reminder interaction logs could not be treated as clean records of task completion or behavioral change. Repeated misses, delayed responses, and low acknowledgment rates often looked meaningful at the data level, but in practice they were shaped by contextual mismatch, interaction barriers, and technical instability. This made reminder logs useful, but also highly vulnerable to misinterpretation if read without household context.

One source of ambiguity was a mismatch between scheduled reminders and how daily life actually unfolded. In Dyad 2, especially, reminders were often set around an ideal routine rather than the PLwD's real timing, which meant that missed reminders did not necessarily indicate disengagement or decline. Tasks were sometimes completed after the active reminder window had expired. As the caregiver explained, "he missed it because the time already left." In the same household, routine interpretation was further complicated by absences from home, transit delays, appointments, and unpredictable return times. These factors made low acknowledgment rates difficult to interpret in isolation, because they could reflect ordinary schedule disruption rather than meaningful change. Reminder systems were therefore not simply capturing whether a task happened, but whether a scheduled prompt successfully intersected with real household life at the right time.

A second source of ambiguity came from the interaction channel itself. Across both dyads, touchscreen use was a major barrier to reliable logging. In Dyad 1, the PLwD explained that "my finger is too big," while the caregiver described the buttons as too close together and difficult to press accurately. In Dyad 2, the caregiver described a broader uncertainty around apparent misses: "you can't tell if something's wrong with the iPad or if… he doesn't want to or if he just didn't click it in time." Technical issues added to this uncertainty, including timing errors after daylight saving time changes, intermittent device problems, and at least one non-functioning bedroom iPad. These problems mattered because they directly shaped what the data appeared to say. A task could be completed but not logged because the response was hard to enter; the reminder disappeared too soon, or the device failed at the wrong moment. The reverse could also happen. In Dyad 1, the caregiver observed that the PLwD could sometimes mark a reminder as done without actually completing the task. In Dyad 2, the caregiver sometimes repaired the record herself after observing that the task had in fact been completed, explaining that she did this because "definitely he did it." Together, these cases show that acknowledgment did not reliably equal completion, and non-acknowledgment did not reliably equal non-completion.

This ambiguity also shaped how caregivers interpreted the reports derived from the logs. In Dyad 2, summary percentages were often hard to make sense of without knowing whether reminders were missed because the household was out, the timing no longer matched the day, or the interaction had failed for some other reason. Week-to-week summaries could therefore appear more meaningful than they really were, especially when routines varied from one week to the next. Caregivers suggested that longer comparison windows and lightweight ways of marking absences or other contextual disruptions would make summaries easier to interpret.

The issue was not simply that the logs were noisy. Rather, they reflected a layered interaction process shaped by scheduling assumptions, accessibility, technical reliability, and caregiver intervention. Their meaning, and the meaning of the summaries built from them, depended on contextual interpretation.

### 4.4   Role-Dependent Usefulness

Because caregiver-facing reports were built from ambiguous interaction data, their usefulness also depended on whether caregivers had the time and context needed to interpret them. Caregiver-facing outputs were not experienced the same way by all caregivers. Their usefulness depended not only on whether a caregiver was remote or in-home, but also on whether they had the time, energy, and capacity to engage with the information.

For remote caregivers, alerts were mainly valuable as reassurance and as prompts for follow-up. In Dyad 1, the caregiver valued notifications because they reduced uncertainty when she was at work and helped her know that the PLwD was "up and active." In Dyad 2, the remote caregiver similarly said that "it is better to get some alert in a sense that I can follow up." In both cases, the value of alerts was not complete explanation, but actionable awareness at a distance.

For the in-home caregiver in Dyad 2, the situation was different. Although reminders were still useful in the moment, she often did not have the capacity to review reports, annotate misses, or engage with feedback tools. She explained, "I don't think I have time to look at that," and later, "I couldn't find time to do it." This was not simply a matter of preference. Her ability to engage with caregiver-facing outputs was constrained by the demands of everyday care, frequent appointments and transit disruptions, and her own health. As she put it, "it's typical for me as a caregiver who's also a patient" and "I'm taking medicine too. Yeah. It's really difficult for me."

At the same time, reports and summaries were not irrelevant to in-home care. In Dyad 2, they were useful in principle, but difficult to sustain and hard to interpret without additional context. Their potential value lay less in frequent review than in offering a broader record of reminder interactions and a way of comparing routines over time. However, week-to-week percentages could look meaningful even when they mainly reflected ordinary disruptions such as appointments, time out of the home, or routine shifts. In this context, longer comparison windows and lightweight ways of marking absences or other contextual disruptions may be more useful than frequent reporting alone.

Caregiver-facing outputs should not be designed as though all caregivers occupy the same position in the care network or have the same ability to act on information. Remote caregivers may benefit from alerts that support reassurance and follow-up, while in-home caregivers may need lower-burden summaries that fit limited time and fragmented attention.

## 5 DISCUSSION

### 5.1 Reminder-based Monitoring as Assistive Infrastructure

Reminder-based monitoring was most useful when understood as part of an assistive caregiving infrastructure rather than a standalone detection tool. Success could not be defined only by independent PLwD response or by the production of a clean behavioral record. Assistive value also came through reassurance, distributed awareness, and support for coordination within the care network.

This broadens how success should be defined for dementia-related assistive systems. Reminder-based awareness should be evaluated not only as a monitoring function, but as a way of supporting the practical and interpretive work caregivers are already doing.

### 5.2 Accessibility and the Data Pipeline

A central contribution of this work is that accessibility is part of the reminder-data pipeline itself, not a separate usability concern. Whether a reminder is recognized as personally relevant, whether the user can physically interact with the interface, whether the timing matches the routine, and whether the device behaves reliably all shape the resulting interaction record. These are not peripheral usability flaws. They directly affect what reminder logs can mean.

Reminder interactions can appear behaviorally meaningful while actually reflecting breakdowns in the assistive interaction itself. Reminder-based monitoring should therefore preserve uncertainty and support contextual interpretation, rather than treat reminder logs as neutral behavioral ground truth. For accessibility research, interpretable outputs and interaction reliability are part of the signal pathway itself.

### 5.3 Design Implications

These findings suggest four concrete design implications, summarized in Table 2. First, reminder-based systems should be designed for shared and household use rather than assuming a single isolated user. In practice, reminders may be noticed, interpreted, or acted on by multiple people. This should be treated as part of normal operation, not only as contamination of the data.

Second, systems should support calibration over time. Reminder schedules, location assignments, and interaction expectations need to be adjusted to real household routines rather than idealized ones. Early deployment data may reflect setup mismatch as much as actual behavior.

Third, systems should preserve uncertainty rather than overclaiming. Reports and alerts should help caregivers interpret patterns but should avoid presenting reminder interaction data as direct evidence of task completion or decline. Lightweight contextual annotation, absence marking, and explicit acknowledgment of ambiguity may be more useful than stronger claims.

Fourth, caregiver-facing outputs should be role-sensitive and easy to personalize. Remote caregivers may benefit from alerts and reassurance, while in-home caregivers may need lower-burden, easier-to-ignore, or more lightweight reporting features. A one-size-fits-all reporting layer is unlikely to fit real care networks.

Table 2: Findings and corresponding design implications

| Empirical finding | Design implication |
|---|---|
| Reminders often functioned through shared household coordination | Design for multi-person and household use |
| Reminder performance depended on routine and placement fit | Support ongoing calibration of schedules and locations |
| Logs were shaped by interaction barriers and technical failures | Preserve uncertainty and avoid overclaiming |
| Output usefulness differed by caregiver role | Personalize alerts and reports by caregiver role |

### 5.4 Limitations and Future Work

This study has several limitations. First, the deployment findings are based on two caregiver-PLwD dyads, so they are context-rich but not broadly generalizable. Second, this paper does not evaluate clinical validity or predictive performance; its focus is on real-world use, interpretability, and assistive value in context. Third, although Remindful was designed within a broader anomaly-aware system architecture, the present paper centers the deployed reminder, logging, alerting, and reporting workflows rather than fully evaluating all technical components of the larger project. Finally, the deployments were relatively short, which limited the actionability of some longer-term summaries and made it harder to assess how calibration and role-specific use might evolve over time.

Future work should examine longer-term home deployments, better support for routine calibration and contextual annotation, and more role-sensitive caregiver interfaces. It should also explore how reminder-based outputs can communicate uncertainty more explicitly without increasing burden.

### 6 CONCLUSION

This paper examined how reminder-based assistive technologies can be extended beyond task prompting to support caregiver awareness in dementia care. The findings show that reminder systems can provide meaningful assistance by reducing uncertainty, supporting coordination, and helping caregivers remain aware of routines over time. However, they also show that reminder interaction logs are shaped by household context, prompt attribution, routine fit, accessibility barriers, and technical reliability.

Reminder-based monitoring should therefore not be treated as a simple detection problem or as a clean behavioral record. Instead, reminder systems should be designed as assistive infrastructures for caregiver interpretation, with outputs that preserve uncertainty and support contextual sensemaking in real homes.